\documentclass[twocolumn,amsmath,amssymb,aps]{aastex61}
\usepackage[fleqn]{amsmath}
\usepackage{mathrsfs}
\usepackage{soul}
\usepackage{xcolor}

\usepackage{ctable}
\usepackage{graphicx}
\usepackage{dcolumn}
\usepackage{bm}
\usepackage{amsmath}
\usepackage{empheq}

\def\Mearth{M_\oplus}

\definecolor{Rust}{HTML}{B7410E}

\received{ }
\revised{ }
\accepted{  }
\submitjournal{The Astrophysical Journal}

\begin{document}

\title{On the stability of low-mass planets with supercritical hydrospheres}
\shorttitle{Hydrospheres of low-mass planets}
\shortauthors{Vivien et al.}
\email{olivier.mousis@lam.fr}



\author{H.~G. Vivien}
\affil{Aix Marseille Univ, CNRS, CNES, LAM, Marseille, France}
\affil{Laboratoire de Physique et Chimie de l'Environnement et de l'Espace (LPC2E), UMR CNRS 7328 - Université d'Orl\'eans, Orl\'eans, France}
\author{A. Aguichine}
\affil{Aix Marseille Univ, CNRS, CNES, LAM, Marseille, France}
\author{O. Mousis}
\affil{Aix Marseille Univ, CNRS, CNES, LAM, Marseille, France}
\author{M. Deleuil}
\affil{Aix Marseille Univ, CNRS, CNES, LAM, Marseille, France}
\author{E. Marcq}
\affil{LATMOS/IPSL, UVSQ Universit\'e Paris-Saclay, Sorbonne Universit\'e, CNRS, Guyancourt, France}
%
%

\begin{abstract}
Short-period and low-mass water-rich planets are subject to strong irradiation from their host star, resulting in hydrospheres in supercritical state. In this context, we explore the role of irradiation on small terrestrial planets that are moderately wet in the low-mass regime (0.2--1$\Mearth$). We investigate their bulk properties for water contents in the 0.01--5\% range by making use of an internal structure model that is coupled to an atmosphere model. This coupling allows us to take into account both the compression of the interior due to the weight of the hydrosphere and the possibility of atmospheric instability in the low-mass regime. We show that even for low masses and low water contents, these planets display inflated atmospheres. For extremely low planetary masses and high irradiation temperatures, we find that steam atmospheres become gravitationally unstable when the ratio $\eta$ of their scale height to planetary radius exceeds a critical value of $\sim 0.1$. This result is supported by observational data, as all currently detected exoplanets exhibit values of $\eta$ smaller than 0.013. Depending on their water content, our results show that highly irradiated and low-mass planets up to $0.9\Mearth$ with significative hydrospheres are not in stable form and should loose their volatile envelope.
\end{abstract}

\keywords{planets and satellites: composition, planets and satellites: ocean, planets and satellites: atmospheres, planets and satellites: detection}

\section{Introduction}
\label{sec:intro}

Earth-like mass planets represent the ultimate goal of exoplanets detection \citep{Ka21}. With today's facilities, their observed number is ever increasing, requiring efforts to enable their characterization, that is accurate measurement of their mass and radius. Due to their small orbital periods, many of those planets are highly irradiated by their parent stars, implying the needs to delevop dedicated interior and atmosphere models to investigate their structure and evolution \citep{Mo20,Ac21,Ag21}. In addition, better understanding of the evolution of the hydrosphere of such planets is key to investigate their potential habitability conditions.

In this study, we investigate the bulk properties of highly irradiated low-mass planets in the 0.2--1$\Mearth$ range, and with water contents varying between 0.01\% and 5\%. These water contents values roughly bracket the water mass fraction (WMF) estimated for the Earth and Jupiter's moon Europa, respectively.

This mass range is also close to the one found by dynamical simulations investigating the formation of water-rich and habitable planets \citep{Ra07}. These mass and water content ranges have not been properly quantified so far in the literature because conventional models \citep[e.g.][]{Ze16,Ze19,Tu19} describe the internal structures of planets with thick atmospheres neglecting the combination of two important effects: i) the compression of the interior due to the weight of the hydrosphere, and ii) the possibility of atmospheric instability at low planetary masses. While the latter can be overlooked for temperate planets, it is critical for high irradiation temperatures and sub-Earth masses. To overcome this issue, we use a self-consistent model in which the atmosphere is coupled with the interior.

\section{Model}\label{sec:methods}

\subsection{Upgrades}\label{ssec:upgrades}

The model used here is built upon the one described in \cite{Ag21}, and includes two major improvements. First, the grid from which the atmospheric properties are interpolated has been extended to surface gravities as low as 1 m.s$^{-2}$, coupled with a revised convergence scheme. In the new convergence scheme, the atmosphere's properties are updated only when the interior model has stabilized. This revision greatly reduces the occurrence of numerical instability in extreme cases, when planetary mass and radius' values are close to the boundaries of the considered range, ensuring convergence to the solution. Second, the atmosphere model now includes a module which assesses the hydrostatic stability of water-rich atmospheres at given planetary mass.
This new and improved version expands the operational validity range towards lower masses and smaller WMFs. The model now allows the computation of atmospheres for planets with gravities ($g_\mathrm{b}$), masses, and boundary temperatures in the ranges $1\leq g_\mathrm{b}\leq 30$ m.s$^{-2}$, $0.2\leq M_\mathrm{p}\leq 20 \Mearth$, and $750\leq T_\mathrm{b}\leq 4500$ K.

\subsection{Principle}\label{ssec:principle}

Our model assumes a planet composed of differentiated layers with various compositions including an iron-dominated core, a lower mantle, an upper mantle and a fluid (from solid to near-plasma) water layer \citep{Mo20,Ag21}. The top of the hydrosphere is a water-dominated atmosphere that follows the prescriptions of \cite{Ma17,Mar19} and \cite{Pl19}.  The structure of the planets assumes hydrostatic equilibrium and adiabatic heat transfer, and takes into account radiative transfer in the atmosphere to generate its physical properties. By doing so, our model self-consistently takes into account the compression of the internal layers of the planet which results from the presence of a hydrosphere.

Three parameters set the distribution of all chemical species in the different layers of the interior model: the fraction of alloy in the core $f_\mathrm{alloy}$, the overall Mg/Si ratio of the planet, and the amount of iron present in the silicate mantle $\mathrm{Mg\#} = \left(\frac{\mathrm{Mg}}{\mathrm{Mg + Fe}}\right)_\mathrm{Mantle}$, which describes the level of differentiation of the planet. Two required compositional inputs of the model are the planet's core mass fraction (CMF) and WMF. Pressure and temperature profiles are integrated from the outside, and require the inputs of the boundary pressure $P_\mathrm{b}$ and the boundary temperature $T_\mathrm{b}$. Finally, the model also requires the input of the planet's mass $M_\mathrm{b}$ (subscript b denotes the mass encapsulated within the boundary of the interior model, and excludes the atmospheric contribution).

Once defined, the input parameters allow for the computation of the planet's internal structure and associated boundary radius. The atmosphere model provides the Outgoing Longwave Radiation (OLR), albedo, thickness and mass of the atmosphere as a function of the bulk mass and radius, and $T_\mathrm{b}$. The atmospheric vertical profile is assumed adiabatic until the  temperature drops to the skin temperature $T_0=T_{\mathrm{eff}}/2^{1/4}$, and isothermal above \citep{Ma17,Mar19}. We choose $P_\mathrm{tr} = 20$ mbar as the pressure of the transiting radius, which corresponds to the altitude where the opacity of the atmosphere is close to unity. Summing the atmosphere's mass and thickness with the bulk ones yields the final mass $M_\mathrm{p}$ and radius $R_\mathrm{p}$ of the planet, respectively. The mass of the water present in the atmosphere is computed by integrating its density profile, and is taken into account in the total water content of the planet. While the mass of the atmosphere is typically much smaller than the planet itself ($\sim 0.01\%$), this correction is crucial for planets with low water mass fractions.

The interior and atmosphere modules are connected at a given boundary pressure $P_\mathrm{b} = 300$ bar, which corresponds to the transition between the vapor and supercritical phases. This threshold is computed for each planet, and thus provides us with an accurate determination of the surface pressure $P_\mathrm{surf}$ exerted on the refractory layers. Similarly, the temperature at the bottom of the atmosphere $T_\mathrm{b}$ is computed iteratively for an input irradiation temperature $T_\mathrm{irr}$, leading to a true surface temperature $T_\mathrm{surf}$.

\subsection{Stability criterion}\label{ssec:stability}

To determine the minimum mass at which a water atmosphere remains stable, we fix a criterion derived from analytical considerations. Modeling the atmosphere beyond $P_\mathrm{tr}$ as a simple theoretical extended isothermal atmosphere\footnote{Gravity variation is taken into account.} of temperature $T_\mathrm{irr}$, the pressure at infinity would be expressed as \citep{Ca17}:

\begin{equation}
P_\infty = P_\mathrm{tr} \exp \left( - \frac{G M_\mathrm{p} m}{k T_\mathrm{irr} R_\mathrm{p}}\right),
\label{eq:p_inf}
\end{equation}

\noindent where $G$ is the gravitational constant, $k$ is the Boltzmann constant, and $m$ is the mass of the molecule composing the atmosphere (here $\mathrm{H_2O}$). This pressure is small but finite, and must correspond to the particle density in the interplanetary medium (IPM), namely $n_\mathrm{IPM} = \frac{P_\mathrm{IPM}}{kT_\mathrm{irr}}$, for the atmosphere to be gravitationally bounded. If the value of $P_{\infty}$ is smaller than $P_\mathrm{IPM}$, then the atmosphere contracts in order to reach the expected IPM pressure, and stabilizes. In the opposite situation, the atmosphere escapes from the planet via the Parker wind mechanism \citep{Wa18}. Combining the inequality $P_\infty > P_\mathrm{IPM}$ with Eq. \ref{eq:p_inf}, we derive a criterion for an atmosphere to be unstable:

\begin{equation} 
H > \eta R_\mathrm{p}, \label{eq:eta}
\end{equation}

\noindent where $H=\frac{kT_\mathrm{irr}R_\mathrm{p}^2}{GM_\mathrm{p} m}$ is the planet's height scale, and $\eta = \left(\ln \frac{P_\mathrm{tr}}{P_\mathrm{IPM}}\right)^{-1}$ is a dimensionless parameter. If $H/R_\mathrm{p}$ is larger than $\eta$, the atmosphere is considered hydrostatically unstable. Because the atmosphere loses its mass by Parker wind in a finite timescale, the value of $\eta$ in Eq. \ref{eq:eta} is in fact underestimated. However computing the timescale of atmospheric loss would require a level of modeling that is beyond the scope of this paper. Instead, we estimate $\eta$ from the output of our model, which can then be used as a conservative estimate of the stability likelihood of a planet.

\subsection{Investigated ranges}\label{ssec:domain_study}

In this study, the compositional parameters $f_\mathrm{alloy}$, Mg/Si, and Mg\# of the different mineral layers are assumed to be equal to 0.13, 1.131, and 0.9, respectively. These values correspond to those derived for the Earth \citep{So07,Br17,Ag21}. To determine the impact of the water content on both the mass-radius relationships and the internal structure of low-mass planets, the WMF of our interior  model has been set to three different values, namely, 0.01\%,  1\%, and 5\%.

The irradiation temperature $T_\mathrm{irr}$ is explored in the 500--2000 K range, which essentially covers the domain of irradiation temperatures of exoplanets discovered so far. Because a change in the core mass fraction is expected to impact both the planet's surface gravity and atmospheric structure, three CMF values have been chosen in our simulations, namely, 0, 0.325, and 0.7. These CMF values correspond to the cases of rocky planets, Earth-like planets \citep{So07}, and Mercury-like planets \citep{St05,Be07}, respectively. Our computation grid follows a logarithmic mass distribution in the 0.2--1.0$\Mearth$ range and linear between 1.0--2.3$\Mearth$, but only planets possessing hydrostatically stable atmospheres are shown. This mass range was previously unavailable as the model from \cite{Ag21} could hardly compute the structure of planets of masses below 0.5 $\Mearth$. Note that the minimum mass of $0.2\Mearth$ has been chosen because it roughly corresponds to the smallest detectable mass today (see Figure.~1 of \cite{De20}).

\begin{figure*}
\centering
\includegraphics[width=1.0\textwidth]{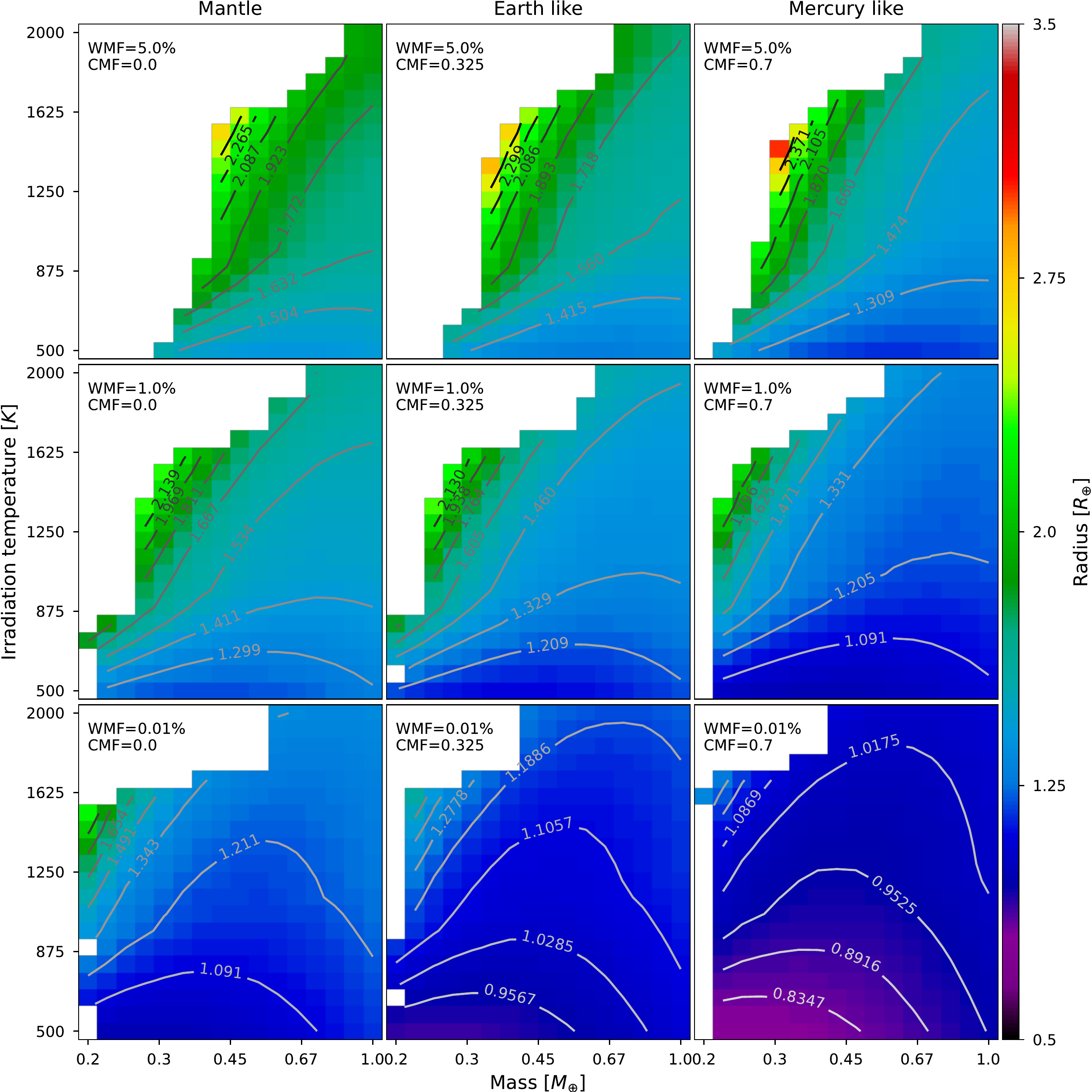}
\caption{Computed planetary radii $R_\mathrm{p}$ at the transiting depth $P_\mathrm{tr} = 20\mathrm{mbar}$ as a function of planetary mass $M_\mathrm{p}$ and irradiation temperature $T_\mathrm{irr}$. Contours show lines of constant planetary radii (in units of Earth radius) to improve the readability of each panel. Left, center and right columns correspond to CMF values that represent pure mantle (0.0), Earth-like (0.325) and Mercury-like (0.7), respectively. Top, middle and bottom rows correspond to WMFs of 5\%, 1\% and 0.01\%, respectively. The missing data correspond to cases excluded from our calculations, due to hydrostatic instability.}
\label{fig:MR_relations}
\end{figure*}

\begin{figure*}
\centering
\includegraphics[width=1.0\textwidth]{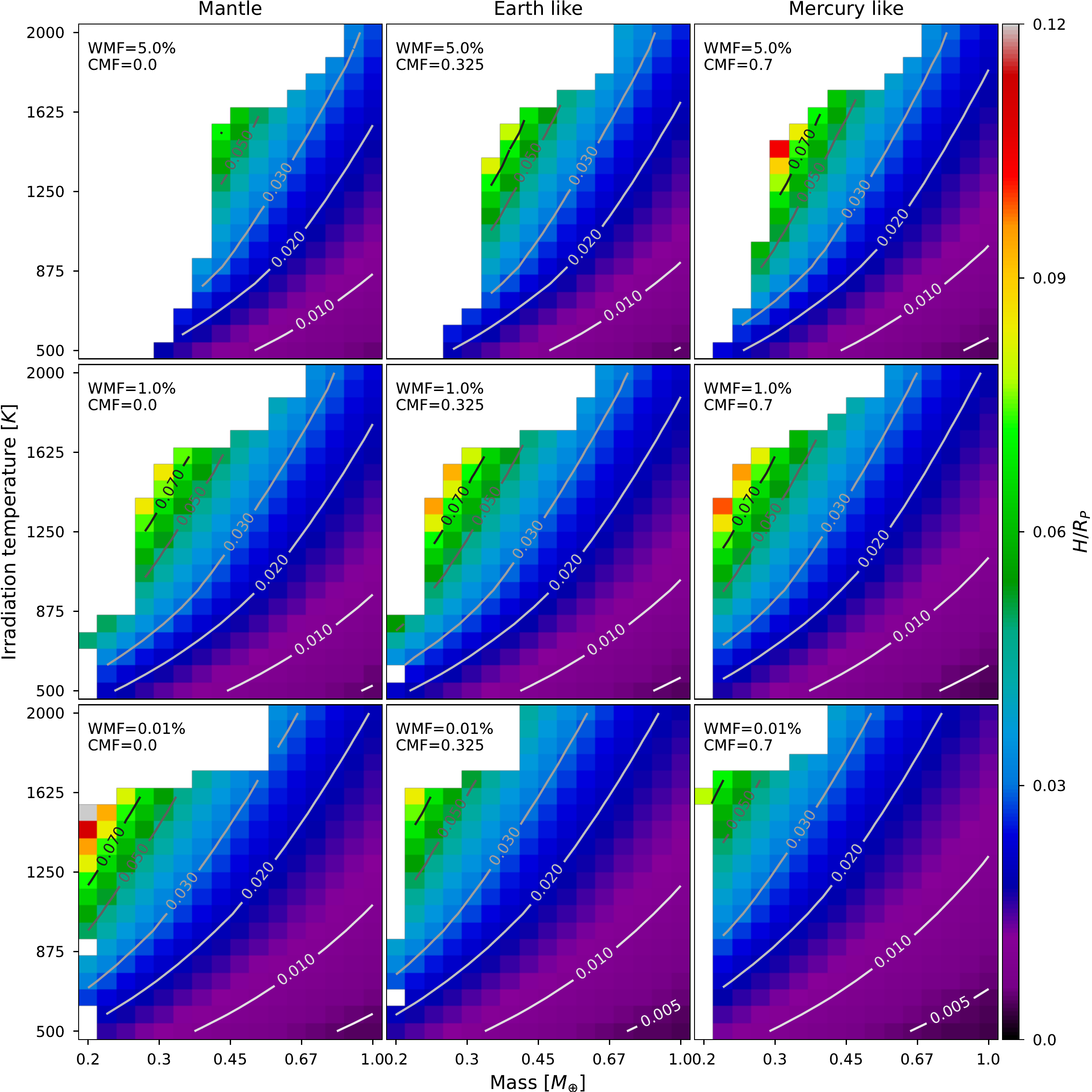}
\caption{Similar to Figure~\ref{fig:MR_relations}, but with a color scheme representing the $H/R_\mathrm{p}$ contours.}
\label{fig:H_RP}
\end{figure*}

\section{Results}\label{sec:results}

Figure~\ref{fig:MR_relations} shows the planetary radii (scaled in Earth radii) as a function of planetary mass and irradiation temperature. This radius corresponds to the altitude where the pressure reaches $P_\mathrm{tr}~=~20~\mathrm{mbar}$ in the hydrosphere. Each column corresponds to CMF values of 0, 0.325 and 0.7, respectively. Each row corresponds to WMF of 0.01\%, 1\% and 5\%, respectively. The different panels of the figure cover a mass range where the atmosphere is in hydrostatic equilibrium. In each panel, the upper left corner exhibit a boundary between hydrostatically stable and unstable planets, which depends on the adopted CMF and WMF. Panels with smaller CMF and higher WMF, which result in lower bulk densities, have more planets with unstable hydrospheres.

In most cases, the planetary radius increases with irradiation temperature and decreases with planetary mass, as an effect of the competition between thermal energy and gravitational binding. This also indicates that the planetary radius is dominated by the atmosphere's thickness. For planets with low water content (i.e. bottom panels of Figure \ref{fig:MR_relations} showing WMF$=0.01\%$), the radius increases with mass when $M_\mathrm{p} \sim 0.5\Mearth$. This indicates a regime where the planetary radius is less sensitive to the atmosphere's thickness. Sharp corners in the boundary between stable and unstable planets (e.g. top panels at $T_\mathrm{irr}\simeq 1500$ K and $M_\mathrm{p}\simeq 0.3~\Mearth$, or middle row at $T_\mathrm{irr}\simeq 1400$ K and $M_\mathrm{p}\simeq 0.25~\Mearth$) are visual artefacts caused by the finite resolution of the grid from which atmospheric properties are interpolated. With an infinite precision, more points would be available between these corners. Therefore, this edge corresponds to cases that are the closest to the physical boundary between stable and unstable atmospheres.

Figure~\ref{fig:H_RP} shows the $H/R_\mathrm{p}$ ratio as a function of planetary mass and irradiation temperature. Contours of constant $H/R_\mathrm{p}$ values seem parallel to the boundary between stable and unstable atmospheres. The highest computed values of $H/R_\mathrm{p}$ indicate that $\eta$ is $\simeq 0.1$ for our model, and has no apparent correlation with the CMF or the WMF. This value is almost $\sim$3--4 times higher than the theoretical estimate ($\eta$ = 0.026--0.035) found for $T_\mathrm{irr}$ = 500--2000 K, $P_\mathrm{tr}$ = 10$^{-3}$--20 mbar, and $n_\mathrm{IPM}$ = 5$\times$ 10$^6$ $\mathrm{m}^{-3}$ \citep{Ca17}. In the currently confirmed population of exoplanets\footnote{http://exoplanet.eu/}, assuming pure H\textsubscript{2}O atmospheres, we find that the $H/R_\mathrm{p}$ ratio is mainly in the $10^{-3}$--$10^{-2}$ range for planets with masses smaller than 2.3 $M_\Earth$. Using Kepler-51~b's planetary parameters derived by \cite{Ma14}, we also find that this planet exhibits one of the highest $H/R_\mathrm{p}$ ratio (0.013$^{+0.006}_{-0.010}$), which is consistent with the aforementioned $10^{-3}$--$10^{-2}$ range. Even if the assumption of a pure steam atmosphere is likely to be unrealistic, this computation gives an upper bound to the criterion. This value is most certainly an observational limitation, better detection capabilities will allow to constrain more precisely the boundary between stable and unstable atmospheres based on the $H/R_\mathrm{p}$ criterion.

A similar criterion was derived by \cite{Ow16} for planets with a pure H/He envelope, based on the ratio between the planet's radius and its Bondi radius ($R_\mathrm{p}/R_\mathrm{B} = 2 H/R_\mathrm{p}$). Interestingly, the authors also find that the atmosphere is lost over a short timescale when this ratio exceeds $0.1$. In addition, \cite{Fo17} did a similar study using the restricted Jeans escape parameter $\Lambda$ = $R_\mathrm{p}/H$ for $\mathrm{H/He}$-rich planets and found that the atmosphere becomes unstable under a value estimated to be $\sim$15--35 (equivalent to the previous criterion $R_\mathrm{p}=0.1R_\mathrm{B}$). Planets with $\Lambda$ values as low as $10$ ($H/R_\mathrm{p}\gtrsim 0.1$) are present in our panels. \cite{Cu17} achieved a similar result and concluded that for the few low-mass planets that exhibit $\Lambda$ values as low as 10, it is likely that either their mass is underestimated or their radius is overestimated. For this latter case, the possible presence of aerosols in the atmosphere would increase its opacity, and lead to an overestimated transit radius \citep{Wa19, Ga20, Oh21}.

\begin{figure}
    \centering
    \includegraphics[width=\linewidth]{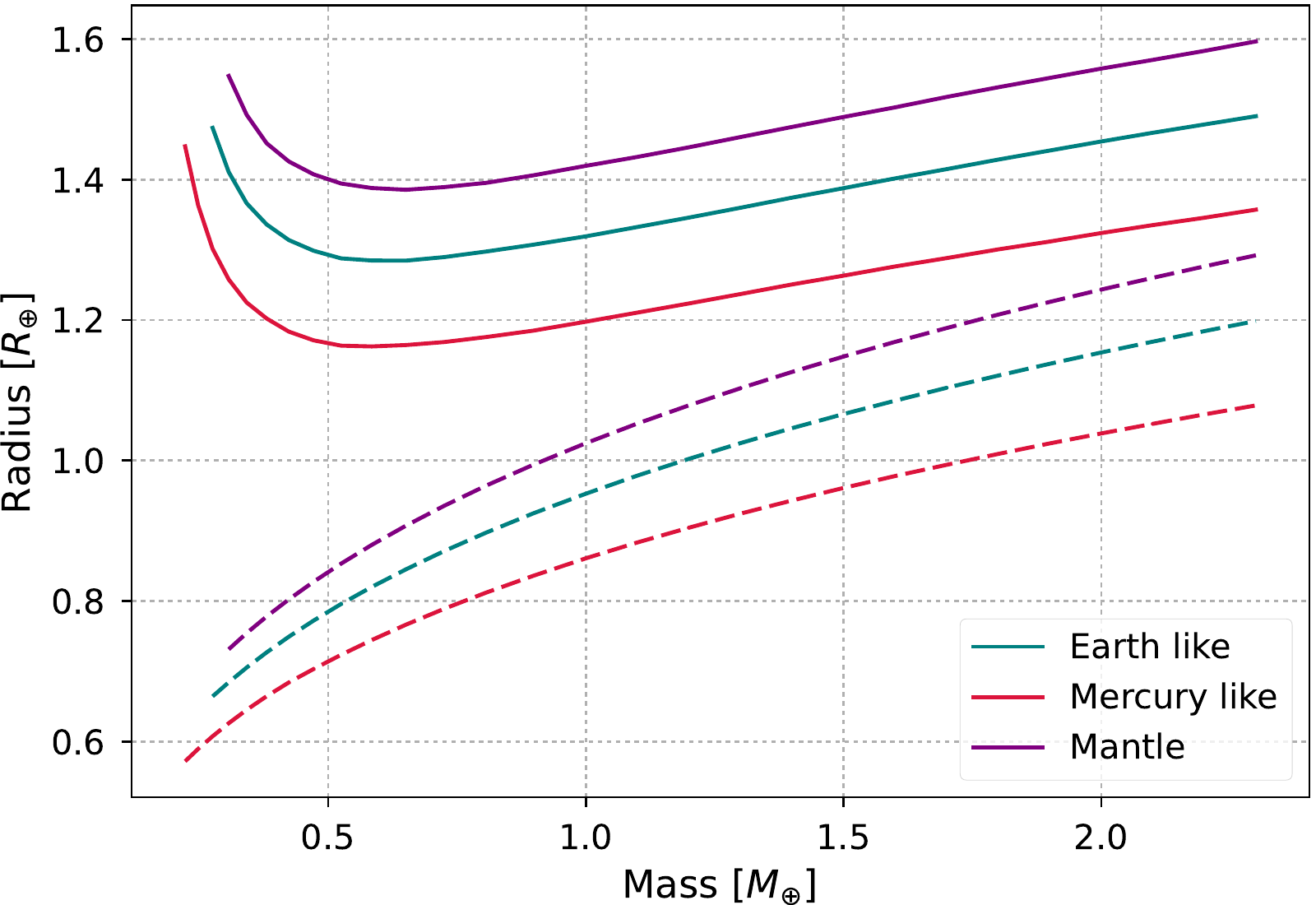}
    \caption{Mass-radius relationships calculated for three CMFs, WMF~=~5\% and $T_\mathrm{irr}~=~500$ K. Dotted and solid lines correspond to the base and the top of the hydrosphere, respectively.}
    \label{fig:cmf_comp}
\end{figure}

Figure~\ref{fig:cmf_comp} shows mass-radius profiles extracted from Figure \ref{fig:MR_relations}, and calculated in the cases of three CMF values, assuming WMF~=~5\% and $T_\mathrm{irr}=500$ K. This figure illustrates the fact that an increase of the CMF increases the surface gravity, and as a consequence reduces the size of the hydrosphere at fixed planetary mass. Even in the case of moderately wet planets (WMF $\sim$5\%), the pressure at the bottom of the hydrosphere can easily reach the GPa range. The subsequent compression of the interior is no longer negligible, and leads to an increase of the surface gravity. This induces thinner atmospheres, for which the blanketing effect is more moderate.

\begin{figure}
    \centering
    \includegraphics[width=\linewidth]{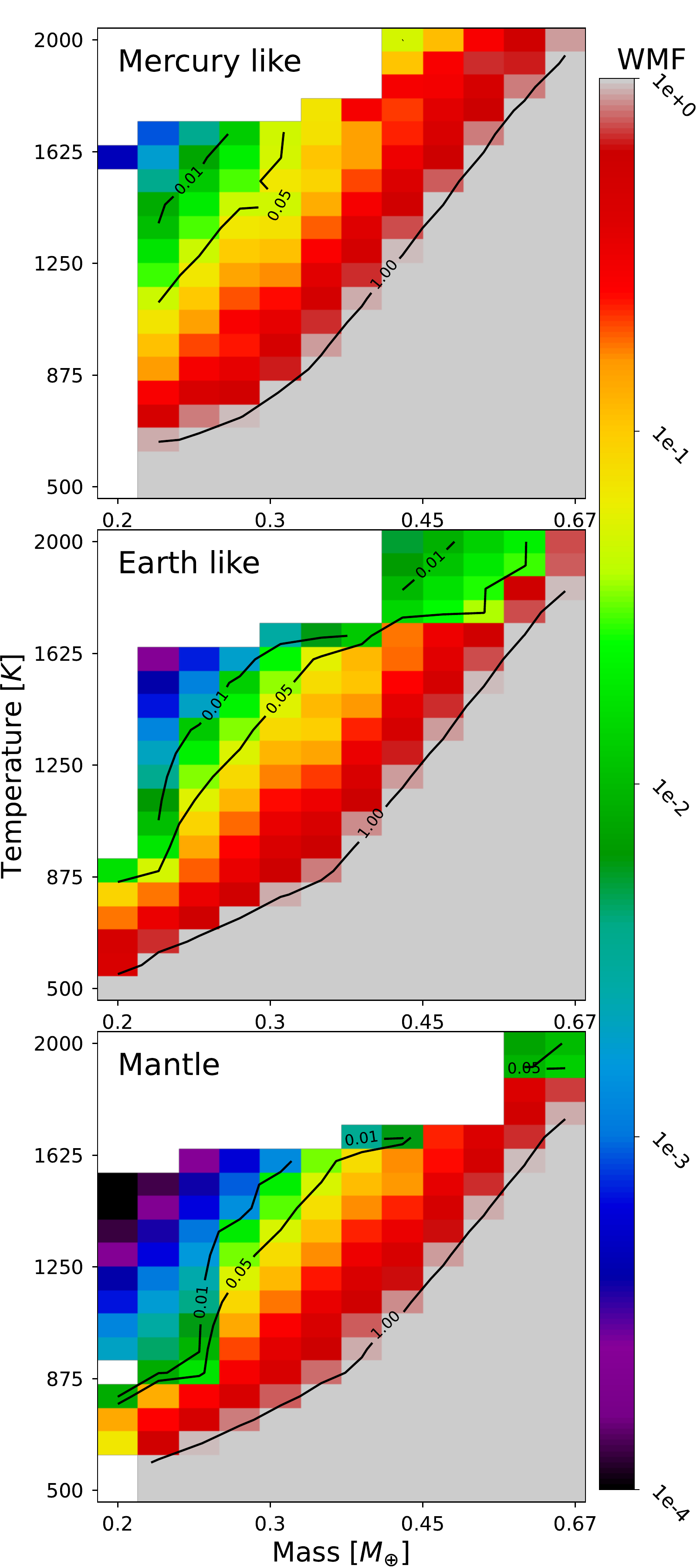}
    \caption{Maximum WMF as a function of planetary mass and irradiation temperature predicted by our model for CMF values representing mantle-like, Earth-like and Mercury-like interiors. All WMF are calculated assuming $H/R_\mathrm{p} = 0.1$, and using a power law fitted to our results. Contours represent different values of WMF. For example, the grey region corresponds to planets fully made of water.}
    \label{fig:max_wmf}
\end{figure}

To illustrate the implications of the $H/R_\mathrm{p}$ boundary criterion, we estimate the maximum WMF as a function of planetary mass and temperature. For each value of $M_\mathrm{p}$ and $T_\mathrm{irr}$ we fit the planetary radius as a function of the WMF with a power-law function to derive a simple relationship between $R_\mathrm{p}$ and the WMF. We then set $H/R_\mathrm{p}=0.1$, and compute the maximal radius using $H=\frac{kT_\mathrm{irr}R_\mathrm{p}^2}{GM_\mathrm{p} m}$. Finally, we invert the fitted radius-to-WMF relation to compute the theoretical maximum WMF that can be achieved. Results are presented in Figure~\ref{fig:max_wmf}, and show that the stability criterion of the atmosphere strongly limits the amount of H\textsubscript{2}O that can be present in low-mass planets.
	

\begin{figure*}[]
\centering
\includegraphics[width=1.0\linewidth]{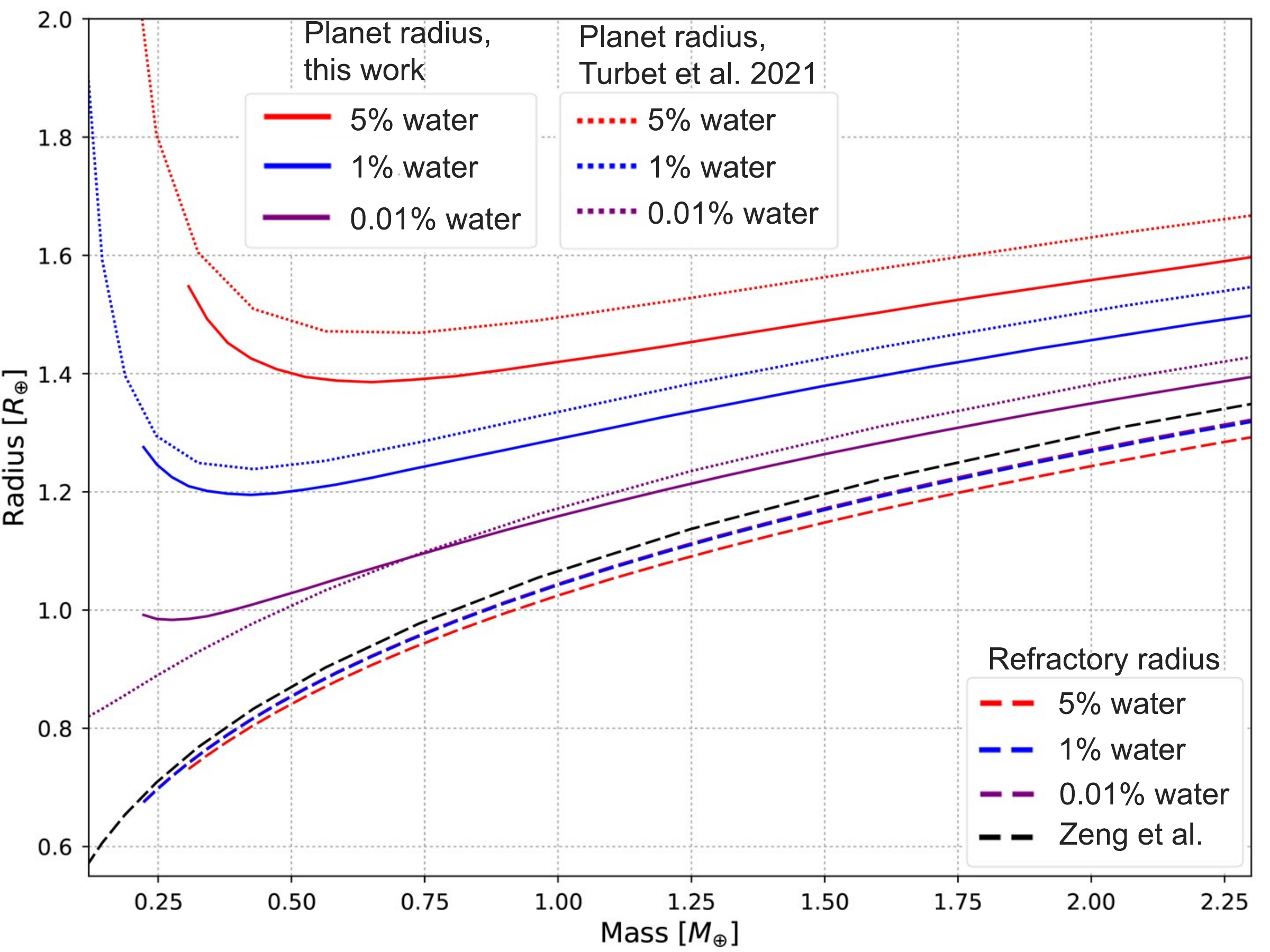}
\caption{Comparison between the mass-radius relationships computed from our model (red, blue, and purple solid lines) and that of \cite{Tu20} (red, blue, and purple dotted lines)
for $T_{\mathrm{irr}}=500\mathrm{K}$ and $\mathrm{CMF} = 0$. Both models define the transiting radius as the altitude where the opacity is close to unity. The colored and black dashed lines correspond to the base of the hydrosphere computed from our approach and that of \cite{Tu20}, respectively.}
\label{fig:comparing}
\end{figure*}

To quantify the feedback of the hydrosphere on the interior and the resulting radius, Figure \ref{fig:comparing} compares our results to those obtained  by \cite{Tu20}. These authors used a pre-defined interior model from \cite{Ze16} coupled with their own atmospheric model. In this comparison, $T_\mathrm{irr}$ is set to $500 \mathrm{K}$, and corresponds to 10.38 times the solar insolation received by the Earth. The mass-radius relationships from \cite{Ze16} are computed at a surface pressure of 1 bar, and become inappropriate for the modeling of wet planets, since the bottom of the hydrosphere can reach up to 10 GPa. This leads to the contraction of the refractory interior of up to 4\% with a WMF of 5\%. In turn, this contraction induces a slight increase of the surface gravity that reduces the atmosphere's thickness. The two effects add up to produce a difference in planetary radii of up to $\sim$10\%. For negligible water contents, both models yield similar results.

When comparing the hydrospheres alone, our model exhibits thinner hydrospheres than the model of \cite{Tu20} in most cases. An additional cause of this could be the use of the steam tables from \cite{Ha84}, which meet several limitations. The corresponding data points can be safely extrapolated up to 3 GPa and 2500 K, but the pressure and temperature at the bottom of the hydrosphere are mostly above this limit (up to $\sim$4500K and 10 GPa). As a result, the adiabatic gradient cannot be extrapolated precisely, leading to an isothermal profile. Under those conditions, the density is overestimated, meaning that the radius is underestimated. In our model, we also use the steam tables from \cite{Ha84} to compute the atmosphere's structure. However, we switch to the equation of state from \cite{Ma19} at $P_\mathrm{b}=300$ bar (supercritical layer). Thanks to this, all equations of state remain in their respective domains of validity within the investigated temperature and pressure ranges. However, given the complexity of both models, it is difficult to list all sources of deviation and precisely quantify their effect. This also highlights the necessity of using identical and up-to-date equations of state when comparing the outcomes of interior models elaborated by different groups.

\section{Conclusion}\label{sec:Discussion}

We have explored the role of irradiation on small terrestrial planets that are moderately wet in the low-mass regime (0.2--2.3$\Mearth$). To that purpose, we have investigated their bulk properties for water contents varying between 0.01$\%$ and 5$\%$ using an upgraded interior and atmospheric structure model based on the one described in \cite{Ag21}. The upgrade of this model includes i) an extension of the grid of atmospheric properties to gravities down to 1 m.s$^{-2}$, and ii) the assessment of the hydrostatic stability of water-rich atmospheres at given planetary mass. This coupling allows us to take into account both the compression of the interior due to the weight of the hydrosphere and the possibility of instability of the water layer in the low-mass regime.

Our results show that the compression of the interior should indeed be taken into account to derive proper planetary structure. In the investigated 0.2--2.3$\Mearth$ mass range, the pressure is found to be in the 0.03--10 GPa range at the bottom of the hydrosphere, depending on the adopted WMF. This causes the refractory part to contract by up to $\sim$4\% in the explored parameter range. The uncompressed case would corresponds to a typical Earth-like atmosphere with a 1-bar pressure at its bottom. Despite this compression, even for low masses and low water contents, the irradiation causes these planets to display inflated atmospheres and increased planetary radius. We also find that the combination of low planetary masses and high irradiation temperatures induce gravitationally unstable steam atmospheres. These planets are subject to fast atmospheric loss when their $H/R_\mathrm{p}$ ratio exceeds $\simeq 0.1$. This value is consistent with the $H/R_\mathrm{p}$ ratios found in the population of detected exoplanets, which are mostly smaller than 0.01. We finally point out that, according to our model, some planets with $H/R_\mathrm{p}$ ratios suggesting rapidly escaping $\mathrm{H/He}$ atmospheres could in fact be water--dominated planets \citep{Cu17,Wa19, Ga20, Oh21}.


Interestingly, the mass limit below which the atmosphere is unstable in our simulations could shift toward larger masses if thermal escape is considered. Given the high irradiation temperatures, thermal escape can lead to a complete loss of the whole hydrosphere \citep{Lo13,Ow13,Ku14,Lo17,Ag21}. This supports the idea that highly irradiated and low-mass planets with thick hydrospheres should not be common, as they lack gravitationally stable atmospheres.

\section{acknowledgements}

O.M. and M.D. acknowledge support from CNES. The project leading to this publication has received funding from the Excellence Initiative of Aix-Marseille Université - A*Midex, a French “Investissements d’Avenir programme” AMX-21-IET-018. We also extend our thanks to the anonymous referee for their insight and useful comments.

\end{document}